   \newcommand{\be}[0]{\begin{equation}}
   \newcommand{\ee}[0]{\end{equation}}
   \newcommand{\ba}[0]{\begin{eqnarray}}
   \newcommand{\ea}[0]{\end{eqnarray}}    
        
\documentstyle[12pt]{article}
\begin{document}
\Large
\hfill\vbox{\hbox{DTP/99/90}
            \hbox{July 1999}}
\nopagebreak

\vspace{0.75cm}
\begin{center}
\LARGE
{\bf Complete Renormalization Group Improvement -
Avoiding Scale Dependence in QCD Predictions}
\vspace{0.6cm}
\Large

C.J. Maxwell

\vspace{0.4cm}
\large
\begin{em}
Centre for Particle Theory, University of Durham\\
South Road, Durham, DH1 3LE, England
\end{em}

\vspace{1.7cm}

\end{center}
\normalsize
\vspace{0.45cm}

\centerline{\bf Abstract}
\vspace{0.3cm}
We show that dimensionful renormalization scheme parameters such
as the renormalization or factorization scale can be
completely eliminated from perturbative QCD predictions
provided that all the ultraviolet logarithms involving
the physical energy scale $Q$ are completely resummed.
\newpage

The problem of the renormalization scale (scheme) dependence
of fixed-order perturbative QCD predictions continues to
frustrate attempts to make reliable determinations of
the underlying dimensional transmutation parameter of the
theory , ${\Lambda}_{QCD}$ (usually ${\Lambda}_{\overline{MS}}$
or ${\alpha}_{s}(M_Z)$ are the fitted quantities). Whilst
a number of proposals for controlling or avoiding this 
difficulty have been advanced \cite{r1,r2,r3,r4} no consensus has been
reached, with the result that in experimental fits attempts
are made to estimate an ad hoc ``renormalization scale''
uncertainty \cite{r5}. This ``theoretical error'' can be
 larger than the actual experimental errors, and in our view
can potentially mislead as to both the central value of
${\Lambda}_{\overline{MS}}$ and the likely importance of
uncalculated higher-order corrections \cite{r4,r6}.\\

It is undeniable that with standard fixed-order renormalization
group (RG-) improvement there {\it is} a scheme dependence
problem. Any proposed ``solution'' must, therefore , amount
to special pleading for a particular choice of scheme, motivated
by albeit reasonable considerations imported from outside
perturbative field theory. Examples include the BLM approach of
Brodsky and collaborators \cite{r3} in which the piece of
the next-to-leading coefficient proportional to the first
beta-function coefficient is absorbed into the coupling,
motivated by various QED examples; or the Principle of
Minimal Sensitivity (PMS) criterion of Stevenson \cite{r1}
, where the scheme is to be chosen such that the perturbative
approximation is as insensitive as possible to changes in scheme.
In this paper we wish to emphasise that the {\it renormalization
scale} dependence of fixed-order QCD perturbation theory is
due to the incomplete nature of the standard RG-improvement
carried out.
The idea will be that the dependence of a dimensionless QCD
observable ${\cal {R}}(Q)$ on the dimensionful parameter $Q$
(e.g. the c.m. energy in ${e}^{+}{e}^{-}$ annihilation) is
obviously completely independent of how the theory is
renormalized. The  perturbative coefficients and the
coupling ${\alpha}_{s}({\mu})$ in contrast manifestly depend
on the dimensionful renormalization scale ${\mu}$ , via the
presence of logarithms ${\ln}({\mu}/{\tilde{\Lambda}})$ , where
$\tilde{\Lambda}$ is universal and depends on the subtraction
procedure used to absorb infinities (usually $\overline{MS}$).
The $Q$-dependence, however , is built by ultraviolet
(UV) logarithms ${\ln}(Q/{{\Lambda}_{\cal{R}}})$ with
${\Lambda}_{\cal{R}}$
completely independent of the renormalization procedure. The
asymptotic $Q$-dependence of ${\cal{R}}(Q)$ is determined by the
dimensionful parameter ${\Lambda}_{\cal{R}}$ which is a physical
property of the observable, independent of the renormalization
scheme \cite{r4}. The key observation is that one should
keep ${\mu}$ {\it independent} of $Q$. If this is done 
standard fixed-order RG-improved predictions are self-evidently
inadequate since they do not satisfy asymptotic freedom
, ${\cal{R}}(Q){\rightarrow}{0}$ as $Q{\rightarrow}{\infty}$.
This property only results if {\it all} the RG-predictable
${\ln}(Q/{{\Lambda}_{\cal{R}}})$ terms are resummed to all-orders
building a $1/{\ln}(Q/{{\Lambda}_{\cal{R}}})$ behaviour. This
summation may be accomplished with any choice of ${\mu}$
, but in doing so all ${\mu}$-dependence cancels between the
${\ln}({\mu}/{\tilde{\Lambda}})$ and ${\alpha}_{s}({\mu})$
terms. Thus the complete resummation of all the RG-predictable
UV logarithms gives ${\mu}$-independent predictions. One has
traded unphysical ${\mu}$-dependence for the correct physical
$Q$-dependence.\\

In the standard RG-improvement one truncates the resummation
of the ${\ln}(Q/{{\Lambda}_{\cal{R}}})$ terms and uses a $Q$-dependent
scale ${\mu}=xQ$, this ``incomplete'' improvement yields
an $x$-dependent result. Clearly the UV logarithms are physical
and leaving out an infinite subset of them gives the resulting
fixed-order approximation an unphysical scale dependence. Of
course there is still residual scheme dependence since fixed-order
predictions will depend on the other dimensionless parameters
specifying the scheme, which can be taken to be the
non-universal beta-function coefficients \cite{r1} , but
dimensionful scheme dependence parameters can always be 
eliminated by complete resummation of all $Q$-dependent
logarithms.\\

There are close links with the Effective Charge approach of
Grunberg \cite{r2} which focusses on building the $Q$-dependence
of ${\cal{R}}(Q)$ . Incomplete improvement with ${\mu}$
chosen to be the effective charge (or fastest apparent convergence
(FAC)) scale is equivalent to the ``complete'' RG-improvement
outlined above. Similar remarks apply to examples such as
moments of leptoproduction structure functions where there
are two (or more) dimensionful scales, for instance the
renormalization scale ${\mu}$ and in addition a factorization
scale $M$. In this case one has ${\ln}(M/{\tilde{\Lambda}})$,
${\ln}({\mu}/{\tilde{\Lambda}})$ logarithms as well as ``physical''
UV logarithms involving Q, which are independent of the
renormalization and factorization conventions. Again all $\mu$
and $M$ dependence is eliminated provided that {\it all} the
physical UV logarithms are resummed \cite{r6a}.\\

We begin by briefly reviewing the problem of parametrizing
RS-dependence, and define the concept of RG-predictable
terms. Consider the dimensionless QCD observable ${\cal{R}}(Q)$
, dependent on the single energy scale $Q$ (we assume massless
quarks). Without loss of generality, by raising   
to a power and scaling, we can arrange that ${\cal{R}}(Q)$
has a perturbation series of the form,
\be
{{\cal{R}}(Q)}=a+{r_1}{a^2}+{r_2}{a^3}+\ldots+{r_n}{a^{n+1}}
+\ldots\;,
\ee
where $a\equiv{\alpha_s}({\mu})/{\pi}$ 
is the RG-improved coupling.
The ${\mu}$-dependence of $a$ is governed by the
beta-function equation,
\be
\frac{{\partial}a}{{\partial}{\ln}{\mu}}   
=-b{a^2}(1+ca+{c_2}{a^2}+{\ldots}+{c_n}{a^n}+{\ldots})\;.
\ee
Here $b$ and $c$ are the first two universal terms of the
QCD beta-function
\ba
b&=&\frac{33-2{N_f}}{6}\;,\\
c&=&\frac{153-19{N_f}}{12b}\;,
\ea
with ${N_f}$ the number of active quark flavours. As
demonstrated by Stevenson \cite{r1} the renormalization scheme
may be completely labelled by the variables 
${\tau}{\equiv}b\ln({\mu}/{\tilde{\Lambda}})$ and the
non-universal beta-function coefficients ${c_2},{c_3},{\ldots}$.
$a({\tau},{c_2},{c_3},{\ldots})$ is obtained as the solution
of the transcendental equation
\be
\frac{1}{a}+c\ln\left(\frac{ca}{1+ca}\right)=
{\tau}-\int^{a}_{0}\;{dx}\;\left(-\frac{1}{B(x)}+
\frac{1}{{x^2}({1+cx})}\right)\;,
\ee
where $B(x)\equiv{x^2}(1+cx+{c_2}{x^2}+{c_3}{x^3}+{\ldots})$.
This is obtained by integrating Eq.(2) with a suitable choice
of boundary condition related to the definition of
${\tilde{\Lambda}}$.\\

For our purposes it will be  useful to label the RS
using $r_1$, the next-to-leading order (NLO) perturbative
coefficient, rather than $\tau$. This is possible because
\cite{r1}
\be
{\tau}-{r_1}={{\rho}_{0}}(Q){\equiv}b\ln(Q/{{\Lambda}_{\cal{R}}})
\;,
\ee
where ${\rho}_{0}$ is an RS-invariant, hence ${\tau}$ can be
traded for ${r_1}$. ${\Lambda}_{\cal{R}}$ is a dimensionful
scale dependent on the particular observable. It is related
to the {\it universal} dimensional transmutation
parameter ${\tilde{\Lambda}}_{\overline{MS}}$ by 
\be
{\Lambda}_{\cal{R}}{\equiv}{e}^{r/b}{\tilde{\Lambda}}_
{\overline{MS}}\;,
\ee
where $r{\equiv}{r}_{1}^{\overline{MS}}({\mu}=Q)$. The
righthand side of Eq.(7) is independent of the subtraction
scheme employed, and as advertised ${\Lambda}_{\cal{R}}$
has a physical significance, being directly related to
the asymptotic $Q$-dependence of ${\cal{R}}(Q)$ \cite{r2,r4,r6}.
Equations (5),(6) can then be used to define 
$a({r_1},{c_2},{c_3},{\ldots})$.\\

We next turn to the RS-dependence of the perturbative
coefficients $r_i$. This must be such as to cancel the
RS-dependence of `$a$' when the series is summed to 
all-orders. The self-consistency of perturbation theory
\cite{r1} demands that the result of a ${\rm{N}}^n$LO 
calculation (terms up to and including ${r_n}{a}^{n+1}$) in
two {\it different} schemes should differ by O(${a}^{n+2}$).
This implies the following dependences of the $r_i$ on the
scheme parameters-
 ${r_2}({r_1},{c_2})$, ${r_3}({r_1},{c_2},
{c_3})$,${\ldots},{r_n}({r_1},{c_2},{c_3},{\ldots},{c_n})$.
The self-consistency requirement can be used to derive
expressions for the partial derivatives of the $r_n$ with
respect to the scheme parameters. For instance for $r_2$ one
has
\be
\frac{{\partial}{r_2}}{{\partial}{r_1}}=2{r_1}+c ,\;\;
\frac{{\partial}{r_2}}{{\partial}{c_2}}=-1 ,\;\;
\frac{{\partial}{r_2}}{{\partial}{c_3}}=0 , \ldots \;\;.
\ee
On integrating these expressions one finds
\ba
{r_2}({r_1},{c_2})&=&{r_1}^{2}+c{r_1}+{X_2}-{c_2}
\nonumber\\
{r_3}({r_1},{c_2},{c_3})&=&{r_1}^{3}+{5\over2}c{r_1}^{2}
+(3{X_2}-2{c_2}){r_1}+{X_3}-{1\over2}{c_3}
\nonumber\\
\vdots & &\vdots\;.
\ea
In general the structure is
\be
{r_n}({r_1},{c_2},{\ldots},{c_n})={{\hat{r}}_{n}}({r_1},{c_2},
{\ldots},{c_{n-1}})+{X_n}-{c_n}/(n-1)\;.
\ee
Here ${\hat{r}}_{n}$ is an $n^{\rm{th}}$ order polynomial in
${r_1}$
which is determined given a complete
${\rm{N}}^{n-1}$LO calculation. $X_n$ is a $Q$-independent
and RS-invariant constant of integration
 and can only be determined given a complete
${\rm{N}}^{n}$LO calculation. ${\hat{r}}_{n}$ is the 
``RG-predictable'' part of $r_n$ , and $X_n$ is 
``RG-unpredictable''.
Thus, given a NNLO calculation in the ${\overline{MS}}$ scheme
with ${\mu}=Q$ one can determine the RS-invariant
\be
{X_2}={{r}_{2}^{\overline{MS}}}({\mu}=Q)
-{({r}_{1}^{\overline{MS}}({\mu}=Q))}^{2}
-c\;{{r}_{1}^{\overline{MS}}({\mu}=Q)}+{c}_{2}^{\overline{MS}}\;,
\ee
where
\be
{c}_{2}^{\overline{MS}}=\frac{77139-15099{N_f}+325{N}_{f}^{2}}
{1728b}\;.
\ee
By a complete ${\rm{N}}^{n}$LO calculation we mean that
${c_2},{c_3},{\ldots},{c_n}$ have been computed as well as
${r_1},{r_2},{\ldots},{r_n}$. We note in passing that
${c}_{3}^{\overline{MS}}$ has now been calculated \cite{r8}.\\

Using Eqs.(9) we can now exhibit the explicit RS-dependence of
the terms of Eq.(1),
\be
{\cal{R}}(Q)=a+{r_1}{a^2}+({r}_{1}^{2}+c{r_1}+{X_2}-{c_2}){a^3}
+({r}_{1}^{3}+{5\over2}c{r}_{1}^{2}+(3{X_2}-2{c_2}){r_1}
+{X_3}-{1\over2}{c_3}){a^4}+{\ldots}\;,
\ee
where $a{\equiv}a({r_1},{c_2},{c_3},{\ldots})$.
We adopt the principle that 
at any given order of Feynman diagram calculation {\it all}
known (RG-predictable) terms should be resummed to all-orders.
Given a NLO calculation ${r_1}$ is known but
${X_2},{X_3},{\ldots}$ are unknown. Thus the complete subset
of known terms in Eq.(13) at NLO is
\be
{a_0}{\equiv}a+{r_1}{a^2}+({r}_{1}^{2}+c{r_1}-{c_2}){a^3}
+({r}_{1}^{3}+{5\over2}c{r}_{1}^{2}-2{c_2}{r_1}-{1\over2}{c_3})
{a^4}+{\ldots}\;.
\ee
The sum of these terms, $a_0$ , can be simply determined using the
following two-step argument. The infinite subset of terms in
Eq.(14) has an RS-independent sum, since the ${X_2},{X_3},{\ldots}$,
-dependent terms cannot cancel their RS-dependence, and we know
that the full sum of Eq.(13) is RS-invariant. Each term is
a multinomial in ${r_1},{c_2},{c_3},{\ldots}$. Using the
RS-independence we can set ${r_1}=0,{c_2}=0,{c_3}=0,{\ldots}$, in
which case all terms but the first in Eq.(14) vanish and we
obtain ${a_0}=a({r_1}=0,{c_2}=0,{c_3}=0,{\ldots},{c_n}=0
,{\ldots})$. So at NLO this corresponds to working in an
``'t Hooft scheme'' with ${c_2}={c_3}={\ldots}=0$ \cite{r9},
and with ${r_1}=0$. From Eq.(6) ${r_1}=0$ corresponds to
${\tau}=b\ln(Q/{{\Lambda}_{\cal{R}}})$ or to an
${\overline{MS}}$ scale ${\mu}={e}^{-r/b}Q$. This is the
so-called ``fastest apparent convergence'' (FAC) or
effective charge (EC) scale \cite{r1,r2}. 
From Eq.(5) we find that ${a_0}$ satisfies
\be
\frac{1}{a_0}+c\ln\left(\frac{c{a_0}}{1+c{a_0}}\right)
=b\ln\left(\frac{Q}{{\Lambda}_{\cal{R}}}\right)\;.
\ee
We note to avoid confusion that the definition of 
$\tilde{\Lambda}$ on which Eq.(5) is based \cite{r1} differs
from that usually used for ${\Lambda}_{\overline{MS}}$. In
terms of the standard definition we have,
\be
{\Lambda}_{\cal{R}}={e}^{r/b}{\left(\frac{2c}{b}\right)}^{-c/b}
{\Lambda}_{\overline{MS}}\;.
\ee

If a NNLO calculation has been completed , then ${X_2}$ can be
determined (as in Eq.(11)), and a further infinite subset of
terms are known and can be resummed to all-orders,
\be
{X_2}{a_0}^{3}={X_2}{a^3}+3{X_2}{r_1}{a^4}+\ldots\;.
\ee
The RS-independence of the sum and the multinomial structure
of the coefficients again leads to a resummed result involving
${a_0}$.
We finally arrive at
\be
{\cal{R}}(Q)={a_0}+{X_2}{a_0}^{3}+{X_3}{a_0}^{4}+{\ldots}
+{X_n}{a_0}^{n+1}+{\ldots}\;,
\ee
which is simply the perturbation series in the RS with
${r_1}={c_2}={c_3}={\ldots}={c_n}={\ldots}=0$.
As is obvious from Eqs.(9), ${X_n}={r_n}({r_1}=0,{c_2}=0,{\ldots},
{c_n}=0)$.\\

Unfortunately the result obtained by resumming all RG-predictable
terms depends on our choice of ${r_1},{c_2},{c_3},{\ldots},{c_n}
,{\ldots}$ as the parameters used to label the scheme. Whilst
this choice is natural and straightforward it is evidently
not unique. We could equally consider a translated set of
parameters-: ${\tilde{r}}_{1}={r_1}-{\overline{r}}_{1} ,\;
{\tilde{c}}_{2}={c_2}-{\overline{c}}_{2} ,\;{\ldots},
{\tilde{c}}_{n}={c_n}-{\overline{c}}_{n}\;,$
where the barred quantities are constants. The partial derivatives
in Eq.(8) with respect to these new parameters are unchanged
so that
\be
\frac{{\partial}{r_2}}{{\partial}{\tilde{r}}_{1}}=2{\tilde{r}}_{1}
+2{\overline{r}}_{1}+c ,\;\; \frac{{\partial}{r_2}}{{\partial}
{\tilde{c}}_{2}}=-1 ,\;\; \frac{{\partial}{r_2}}{{\partial}
{\tilde{c}}_{3}}=0\;\;,   
\ee
which on integration yields
\ba
{r_2}({\tilde{r}}_{1},{\tilde{c}}_{2})&=& {{\tilde{r}}_{1}}^{2} +
2{\overline{r}}_{1}{\tilde{r}}_{1}+c{\tilde{r}}_{1}+
{\tilde{X}}_{2}-{\tilde{c}}_{2} \nonumber\\
\vdots& &\vdots
\ea
with general structure
\be
{r_n}({\tilde{r}}_{1},{\tilde{c}}_{2},{\ldots},{\tilde{c}}_{n})=
{\hat{r}}_{n}({\tilde{r}}_{1},{\tilde{c}}_{2},{\ldots},{\tilde{c}}_{n-1}
)+{\tilde{X}}_{n}-{\tilde{c}}_{n}/({n-1})\;.
\ee
The ${\tilde{X}}_{n}$ are again constants of integration which
are unknown unless a complete ${\rm{N}}^{n}$LO calculation has
been performed. If one applies the same rationale as before, where
all RG-predictable terms are to be resummed, one finds , analogous
to Eq.(18),
\be
{\cal{R}}(Q)={\overline{a}}+{\overline{r}}_{1}{{\overline{a}}}^{2}
+{\tilde{X}}_{2}{\overline{a}}^{3}+{\ldots}+{\tilde{X}}_{n}
{\overline{a}}^{n+1}+{\ldots}\;,
\ee
with ${\overline{a}}{\equiv}a({r_1}={\overline{r}}_{1},
{c_2}={\overline{c}}_{2},{\ldots},{c_n}={\overline{c}}_{n},
{\ldots})$, which is just the perturbation series in the scheme
with ${r_1}={\overline{r}}_{1},{c_2}={\overline{c}}_{2},
{\ldots},{c_n}={\overline{c}}_{n},{\ldots}$, or equivalently
with ${\tilde{r}}_{1}=0,{\tilde{c}}_{2}=0,{\ldots},{\tilde{c}}_{n}
=0,{\ldots}$.\\
                                       
Thus by itself the principle of resumming all the RG-predictable
terms at any given order of Feynman diagram calculation does
not abolish the scheme dependence problem since the subset of
RG-predictable terms depends on how the scheme dependence is
parametrized, the parametrization ambiguity being precisely
equivalent to the scheme dependence ambiguity. However, as we
shall now argue, the parameter $r_1$ has a special status, being
connected with the dimensionful renormalization scale $\mu$
and the physical energy scale $Q$ on which ${\cal{R}}(Q)$
depends. Thus from Eq.(6) we see that we may write
\be
{r_1}({\mu})=\left(b\ln\frac{{\mu}}{{\tilde{\Lambda}}_
{\overline{MS}}}-b\ln\frac{Q}{{\Lambda}_{\cal{R}}}\right)\;,
\ee
with ${\mu}$ taken to be the $\overline{MS}$ scale as is
customary. To simplify the discussion let us temporarily set
$c=0$ and work in a scheme with ${c_2}={c_3}={\ldots}={c_n}
={\ldots}=0$. Then the coupling $a({\mu})$ is given by
\be
a({\mu})=1/b\ln({\mu}/{{\tilde{\Lambda}}_{\overline{MS}}})\;,
\ee
and the NLO RG-improvement in Eq.(14) becomes a geometrical
progression in $r_1$,
\be
{\cal{R}}(Q){\approx}a({\mu})+{r_1}({\mu}){a^2}({\mu})
+{r}_{1}^{2}({\mu})
{a}^{3}({\mu})+{\ldots}+{r}_{1}^{n}{a}^{n+1}({\mu})+{\ldots}\;.
\ee
Substituting Eq.(23) for ${r_1}({\mu})$, summing the geometrical
progression and using Eq.(24) for $a({\mu})$ yields
\be
{\cal{R}}(Q){\approx}a({\mu})/\left[1-\left(b\ln\frac{\mu}
{{\tilde{\Lambda}}_{\overline{MS}}}-b\ln\frac{Q}{{\Lambda}_
{\cal{R}}}\right)a({\mu})\right]=1/b\ln(Q/{\Lambda}_
{\cal{R}})\;.
\ee
We see explicitly the cancellation of the unphysical
${\ln}({\mu}/{\tilde{\Lambda}}_{\overline{MS}})$ terms in
${r_1}({\mu})$ with those in $a({\mu})$ to generate the
correct physical $Q$-dependence of ${\cal{R}}(Q)$,
\be
{\cal{R}}(Q){\approx}1/b{\ln}(Q/{\Lambda}_{\cal{R}})+
O{(1/b{\ln}(Q/{\Lambda}_{\cal{R}}))}^{3}\;.
\ee
In contrast with standard NLO RG-improvement one has
\be
{\cal{R}}(Q){\approx}a({\mu})+{r_1}({\mu}){a}^{2}({\mu})\;,
\ee
where ${\mu}=xQ$ is taken proportional to $Q$ with $x$ a
dimensionless constant, and $x=1$ the so-called ``physical
scale'' is often favoured. The resulting $Q$-dependence is
\be
{\cal{R}}(Q){\approx}1/b({\ln}(xQ/{\tilde{\Lambda}}_{\overline
{MS}})+O{(1/b{\ln}(xQ/{\tilde{\Lambda}}_{\overline{MS}}))}^{2}\;,
\ee
which is, of course, $x$-dependent. All $x$-dependence cancels
and the physical $Q$-dependence of Eq.(27) results if the
geometric progression of Eq.(25) is not truncated but is
resummed to all-orders.
Truncating the resummation can result in considerable error
in the extraction of ${\Lambda}_{\overline{MS}}$ from comparisons
of NLO perturbative results with experiment. For many
${e}^{+}{e}^{-}$ jet observables one has 
$r={r}_{1}^{\overline{MS}}({\mu}=Q){\approx}10$ \cite{r4}
and the truncation of a geometric progression with a common
ratio $ra{\approx}1/2$ (taking $a={a}_{\overline{MS}}({\mu}
={M_Z}){\approx}0.05$) can lead to a sizeable overestimate
of the true ${\Lambda}_{\overline{MS}}$ unless fortuitously
the, as yet unknown , NNLO and higher invariants ${X_2},{X_3},
{\ldots}$ compensate.\\

The complete resummation of Eq.(25) is forced on one if the
renormalization scale ${\mu}$ is kept {\it independent} of
the physical energy scale $Q$ (i.e. $\mu$ held constant) since
then the ${\ln}(Q/{\Lambda}_{\cal{R}})$ UV logarithms are the
only source of $Q$-dependence. Standard fixed-order NLO
RG-improvement is then manifestly unsatisfactory since it
does not satisfy asymptotic freedom, ${\cal{R}}(Q){\rightarrow}
0$ as $Q{\rightarrow}{\infty}$. Instead
\be
{\cal{R}}_{NLO}(Q)=a(\mu)+\left(b{\ln}\frac{\mu}{{\tilde{\Lambda}
}_{\overline{MS}}}-b{\ln}\frac{Q}{{\Lambda}_{\cal{R}}}\right)
{a^2}(\mu)\;,
\ee
tends to $-{\infty}$ as $Q{\rightarrow}{\infty}$ with
$\mu$ fixed. Only if the complete resummation in Eq.(25) is
carried out does one build the correct $1/{\ln}(Q/{\Lambda}_
{\cal{R}})$ behaviour.\\

Notice that use of the parametrization with ${\tilde{r}}_{1}=
{r_1}-{\overline{r}}_{1}$ and ${\overline{r}}_{1}=
b{\ln}(xQ/{\tilde{\Lambda}}_{\overline{MS}})-b{\ln}(Q/
{\Lambda}_{\cal{R}})$ is equivalent to standard NLO RG-improvement
with ${\mu}=xQ$ and
yields the $x$-dependent result in
Eq.(29) when all the NLO RG-predictable ${\tilde{r}}_{1}$
terms are resummed. The resummation is then manifestly
incomplete with respect to UV logarithms since the
${r_1}a^{2}$ term is split into ${\tilde{r}}_{1}a^{2}+
{\overline{r}}_{1}a^{2}$ with the ${\overline{r}}_{1}{a}^{2}$ 
term not summed. Furthermore the constants of integration
${\tilde{X}}_{2},{\tilde{X}}_{3},{\ldots},$ which are unknown
and hence omitted at NLO now contain ${\ln}(xQ/{\tilde{\Lambda}}_
{\overline{MS}})$ and ${\ln}(Q/{\Lambda}_{\cal{R}})$ terms.
For instance 
\be
{\tilde{X}}_{2}={X_2}+{\left(b{\ln}\frac{xQ}{{\tilde{\Lambda}}_
{\overline{MS}}}-b{\ln}\frac{Q}{{\Lambda}_{\cal{R}}}\right)}^{2}
\;.
\ee
If these terms are included in the resummation the $x$-dependence
disappears order-by-order reproducing the complete resummation
of Eq.(25) and giving the correct physical $Q$-dependence in
Eq.(27).
So the parameter $r_1$ has a special            
status and should not be translated if one wishes to
completely resum {\it all} the UV logarithms and reconstruct the
correct physical $Q$-dependence.\\

With realistic non-zero $c$ the NLO complete RG-improvement
(CORGI) of Eq.(14) sums to $a_0$ satisfying Eq.(15) . This
may be written in closed form as
\ba
{\cal{R}}(Q)&{\approx}&\frac{-1}{c[1+W(z(Q))]}\nonumber\\
z(Q)&{\equiv}&-\frac{1}{e}{\left(\frac{Q}{{\Lambda}_{\cal{R}}}
\right)}^{-b/c}\;,
\ea
where $W(z)$ is the Lambert $W$-function defined implicitly by
$W(z){\exp}(W(z))=z$ \cite{r7}.
This is of course equivalent to standard ``incomplete'' NLO
RG-improvement with ${\mu}={e}^{-r/b}Q$, the FAC (or EC) scale
, as noted earlier. Whilst CORGI yields ${\mu}$-independent
results there is still a dependence on the other dimensionless
scheme parameters ${c_2},{c_3},{\ldots}$. That is one should
parametrize the RS-dependence using $r_1$ (i.e. ${\overline{r}}_{1}
=0$), but there is no preference for any particular
parametrization ${\tilde{c}}_{2},{\tilde{c}}_{3},{\ldots}$.
In the effective charge approach of Grunberg \cite{r2}
one chooses ${\overline{c}}_{2},{\overline{c}}_{3},{\ldots},
{\overline{c}}_{n}$ so that the ${\tilde{X}}_{2},{\tilde{X}}_{3},
{\ldots},{\tilde{X}}_{n}$ are zero at ${\rm N}^{n}$LO, 
corresponding to ${r_1}=0,{r_2}=0,{\ldots},{r_n}=0$.
The use of ${c_2},{c_3},{\ldots},{c_n}$ as parameters corresponds
to an 't Hooft scheme \cite{r9} ${\overline{c}}_{2}=0,
{\overline{c}}_{3}=0,{\ldots},{\overline{c}}_{n}=0$, and
is {\it a priori} equally reasonable \cite{r2}. At NLO
one conventionally makes this choice in any case.\\

We finally note that inverting the relation of Eq.(32)
leads to the property of asymptotic scaling \cite{r11}
\be
\lim_{Q{\rightarrow}{\infty}}Q{\cal{F}}({\cal{R}}(Q))=
{\Lambda}_{\cal{R}}\;,
\ee
where ${\cal{F}}(x)$ is the universal QCD scaling function
\be
{\cal{F}}(x){\equiv}{e}^{-1/bx}{(1+1/cx)}^{c/b}\;.
\ee
Exactly this property is used in lattice gauge theory
calculations of lattice coupling to assess how close to
the continuum limit of infinite inverse lattice spacing
one is. By using the exact relation between
${\Lambda}_{\cal{R}}$ and the {\it universal} parameter
${\Lambda}_{\overline{MS}}$ in Eq.(16) one arrives at
the {\it universal} asymptotic scaling relation
\be
\lim_{Q{\rightarrow}{\infty}}Q{\cal{F}}({\cal{R}}(Q))
{e}^{-r/b}{(2c/b)}^{c/b}={\Lambda}_{\overline{MS}}\;.
\ee
This relation can be used analogously to the lattice
scaling relation to assess how close to asymptotia in
$Q$ one is at current energies, for QCD observables calculated
to NLO. One simply inserts the data for ${\cal{R}}(Q)$ , and
the corresponding NLO corrections $r{\equiv}{r}_{1}^
{\overline{MS}}({\mu}=Q)$ , into Eq.(35). The asymptotic
prediction is that all the data should lie on a single
horizontal straight line corresponding to ${\Lambda}_
{\overline{MS}}$. Deviations from this unambiguously
indicate the presence of sub-asymptotic effects , and
enable the estimation of relative differences between the
uncalculated NNLO invariants $X_2$ , and possible
power corrections, for different observables. Observation of  
the scaling property at some approximate level provides
a well-founded starting point for further analyses attempting
to resum large logarithmic corrections for jet observables
or predict power corrections. 
In particular the sub-asymptotic effects are contained in
a non-universal factor ${\cal{G}}({\cal{R}}(Q))$ which
approaches unity as $Q{\rightarrow}{\infty}$, so that
\be
Q{\cal{F}}({\cal{R}}(Q)){\cal{G}}({\cal{R}}(Q)){e}^{-r/b}
{(2c/b)}^{c/b}={\Lambda}_{\overline{MS}}\;
,
\ee
with
\be
{\cal{G}}({\cal{R}}(Q))=1-({X_2}/b){\cal{R}}(Q)+{\ldots}+
({K_0}/{\cal{R}}^{2}){\cal{R}}^{-c/b}{e}^{-1/b{\cal{R}}}+
{\ldots} \;.
\ee
The $K_0$ term represents a possible power correction, here taken
to be $1/Q$. Direct fits of $X_2$ and $K_0$ to the
$Q$-dependence of the data can be performed along the lines
discussed in Ref.[6].\\

This is to be contrasted
with standard experimental analyses \cite{r5} which attempt to
assess the
size of uncalculated higher-order   corrections by variation
of the chosen renormalization scale ${\mu}=xQ$. As we have
attempted to indicate here such an approach can be
misleading , and leads to no information on the size of
the uncalculated  higher-order invariants ${X_2},{X_3},{\ldots}$.

\section*{Acknowledgements}

We would like to thank John Campbell and Nigel Glover
for a productive collaboration and numerous discussions
which were instrumental in developing the ideas reported
here.

\end{document}